\documentclass{proc}
\usepackage{stfloats}

\usepackage{amsfonts}
\usepackage{amsmath}
\usepackage[pdftex]{graphicx}
\usepackage[utf8]{inputenc}
\usepackage{caption}
\usepackage{subcaption}
\usepackage{epstopdf}
\usepackage{color}
\usepackage{cite}
\input{tcilatex}
\begin{document}

\title{Predicting Regional Economic Indices Using Big Data\\ Of Individual Bank Card Transactions}
\author{Stanislav Sobolevsky$^{1}$, Emanuele Massaro$^{1}$, Iva Bojic$^{1}$, Juan Murillo Arias$^{2}$, Carlo Ratti$^{1}$ \\
\\
{$ ^{1}$}Senseable City Lab, Massachusetts Institute of Technology, Cambridge, MA, US \\
{$ ^{2}$}New Technologies, Banco Bilbao Vizcaya Argentaria, Madrid, Spain \\
stanly@mit.edu, emassaro@mit.edu, ivabojic@mit.edu, juan.murillo.arias@bbva.com, ratti@mit.edu}
\maketitle

\begin{abstract}
For centuries quality of life was a subject of studies across different disciplines. However, only with the emergence of a digital era, it became possible to investigate this topic on a larger scale. Over time it became clear that quality of life not only depends on one, but on three relatively different parameters: social, economic and well-being measures. In this study we focus only on the first two, since the last one is often very subjective and consequently hard to measure. Using a complete set of bank card transactions recorded by Banco Bilbao Vizcaya Argentaria (BBVA) during 2011 in Spain, we first create a feature space by defining various meaningful characteristics of a particular area performance through activity of its businesses, residents and visitors. We then evaluate those quantities by considering available official statistics for Spanish provinces (e.g., housing prices, unemployment rate, life expectancy) and investigate whether they can be predicted based on our feature space. For the purpose of prediction, our study proposes a supervised machine learning approach. Our finding is that there is a clear correlation between individual spending behavior and official socioeconomic indexes denoting quality of life. Moreover, we believe that this \emph{modus operandi} is useful to understand, predict and analyze the impact of human activity on the wellness of our society on scales for which there is no consistent official statistics available (e.g., cities and towns, districts or smaller neighborhoods).
\end{abstract}

\section{Introduction}

For centuries great thinkers discussed the essential qualities of good societies and of a good life. In the end three major philosophical approaches were proposed to determine quality of life \cite{Brock_1989}: dictated by normative ideals based on religious, philosophical or other systems; based on whether people can obtain the things they desire; and described by the way people experience their lives. Those three philosophical approaches were then "translated" into three measurable indices denoting quality of life: social, economic and subjective well-being measures. 

However, not all parameters are seen equally important for different parties. For example, policy makers put more emphases on economic ones, while some others think that individuals experience their lives subjectively and that economic parameters can sometimes be negatively correlated with certain quality of life facets such as people leisure time or their need to live in healthy environments \cite{diener1997measuring, roser2014happiness, clark2011will}. In contrast to economic parameters that measure one's ability to obtain the marketplace goods and services they choose, social indices question economic growth in terms of whether more is always better \cite{palys1980social, knutsen2008gdp}. Finally, well-being research focuses on one's conscious experiences described by hedonic feelings or cognitive satisfactions \cite{berridge1996food, bergheim2006measures}.

The goal of our research is to provide a model that can be used to predict quality of life on a city level. Currently different quality of life parameters are calculated on a much coarse-grained scale than cities (e.g., for regions, provinces or the whole countries). Our motivation is to be able to further "zoom in" to the city or neighborhood level providing people with indices that can describe quality of life for the exact location where they live, rather than provide them with average values that very often have large deviations. This information can be then used either by the city policy makers or people deciding where to live \cite{florida2010s}.

Without going into the discussion of which quality of life parameters are more important, in this paper we choose to use three social and three economic quantities. We left out subjective well-being factors as it is hard to find a reliable and consistent country-wide official statistic source for them, which makes this problem a subject for a separate study. Out of a very wide variety of social and economic indices, we included major quantities such as Gross Domestic Product (GDP), housing price level, unemployment rate, as well as social ones such as crime rate, percentage of higher education and life expectancy. 

Namely, GDP is a traditional outcome metric of developmental economics that measures aggregated economic activity within a given country, housing prices are important parts of capital markets in which money is provided for periods longer than a year, while labor markets that function through the interaction of workers and employers are to a great extent characterized by unemployment rates. Moreover, social parameters denote how safe people should feel living in a certain area, what the level of education for that area is and how long people are expected to live. All of which are important when accessing their quality of life.

In order to develop a model that is able to predict quality of life factors for various locations at different scales, we propose a three step process in which we first define and calculate different microeconomics indicators using individual bank card transaction records collected by Banco Bilbao Vizcaya Argentaria (BBVA), then use those microeconomics indicators as inputs of machine learning process to teach our model how to predict six quality of life parameters on Spanish province level for which we have reliable official statistical data, and then finally use our model to predict them on a much more fine-grained spatial scale for which the official statistical data is inconsistent or non-existent. In this paper we describe the first two steps of the process, while the third one is a subject for future work. 
 
The rest of the paper is organized as follows. In Section \ref{rw} we give a short overview of related work starting with describing studies made before the digital era and ending with studies based on bank card data. Section \ref{dataset} provides insights into dataset that we use in this paper -- BBVA dataset of bank card transactions preformed in the whole Spain. In Section \ref{methodology} we describe the methodology that we used to conduct research for this paper together with all technical aspects of our modeling procedure. Finally, in Section \ref{conclusions} we discuss the results presented in Section \ref{res}.

\section{Related work}
\label{rw}

In the era where the usage of digital technologies is so omnipresent, people every day leave more digital trails than we are currently able to process. No matter whether they surf the Internet, post on their social media, twit, publish their geotagged photographs or use their bank cards to make their purchases, people create their own digital footprints. In extensive amount of related work scholars used such datasets for different research purposes such as for studying human dynamics through cell phone data \cite{ratti2006mlu, gonzalez2008uih, hoteit2014estimating, amini2014impact, kung2014exploring}, social media posts \cite{hawelka2014geo, paldino2015urban, podobnik2012web, smailovic2014calculating, podobnik2013analysis} or vehicle GPS traces \cite{santi2014quantifying, kang2013exploring}. Results of those studies can be also used for a variety of applications including support of decision-making in regional \cite{Ratti2010redrawing, Sobolevsky2013delineating} or urban land use planning \cite{pei2014new, grauwin2014towards}.

In this paper we use BBVA bank card transaction dataset to learn more about people quality of life. It is not only important how much or on what people spend their money, but also a broad variety of other more specific characteristic does matter as it is explained in Section \ref{methodology}. In recent years similar datasets have already been used to investigate people individual spending patterns, but to the best of our knowledge, nobody used these microeconomic trends to calculate quality of life parameters. In the rest of this section we will sum up related research.

People individual spending activities, which were investigated before the digital era, collected data using field studies \cite{lloyd54si}, questionnaires \cite{childers2001}, surveys from users \cite{dholakia1999} or retailers \cite{buckinx2005}. The focus of those studies was mostly on finding correlations between demographic factors (e.g., age group, gender, education level, occupation or income) and either shopping patterns \cite{dholakia1999, bhantagar2000, hui2007} or predisposition to use different payment methods such as bank cards, checks or money \cite{boeschoten1998, hayhoe2000, bounie2006, borzekowski2008}. The results of studies investigating the latter correlations were inconclusive in both cases of gender and age groups. Namely, some studies concluded that women are more prone to use bank cards for their purchases \cite{hayhoe2000, borzekowski2008}, while other pointed towards their preference for checks over cash or cards \cite{bounie2006}. Moreover, in some studies age is reported to lower the probability of card usage \cite{borzekowski2008}, while others reported no significant effect \cite{bounie2006}.

Since the aforementioned findings were mostly based on survey results, they may have been affected by the fact that people could have altered their answers knowing that they were "monitored". Today in the digital era in some cases information about people behavior is collected even without them being aware of that, let alone with their informed consent. However, as bank card transaction data is highly sensitive and includes a lot of private information, access to it has been so far highly restricted. Therefore, related studies have been mostly focused on card systems \cite{chan1999frauddetection, rysman2007, mahmoudi2015detecting}, rather than on human behavior that can be derived from people using them. 

In a few studies that do focus on extracting some features of human behavior based on their bank card transaction records, scholars investigate how individual spending is affecting those individuals. For example, some studies wanted to uncover the predictability of people spending choices and their relationship to their wealth \cite{krumme2013patterns} or examine the relationship between wealth/income/financial literacy and the failure to make the minimum monthly payment on their credit cards even when having enough funds on their deposit accounts to make the required payment \cite{scholnick2013impact}. 

In addition to that, our previous studies shed some light on how macroeconomic patterns emerge from microeconomics ones.
Namely, in \cite{sobolevsky2014mining, sobolevsky2015cities} using the same BBVA dataset as we are using in this paper, we presented city classification based on customer individual behavior that could not have been observed from the official Spanish socioeconomic statistics. Moreover, recently we utilized this dataset for showing how behavior or foreign visitors depends on their country of origin \cite{sobolevsky2014money} and the city size \cite{sobolevsky2015scaling}. This gives an idea of possible correlation between individual spending patterns and quality of life parameters in the city. The closest related work to this idea is presented in \cite{shen2014credit} and \cite{sabatini2011can}. The former one shows how a relationship between people debt and their psychological well-being evolves over time, while the latter one finds a positive correlation between subjective well-being and e-shopping. Although both studies investigate quality of life well-being factors, they are doing it on an individual level rather than on a city level as proposed in this paper.

\section{Dataset}
\label{dataset}

\begin{figure}[b!]
  \centering
      \includegraphics[width=0.88\columnwidth]{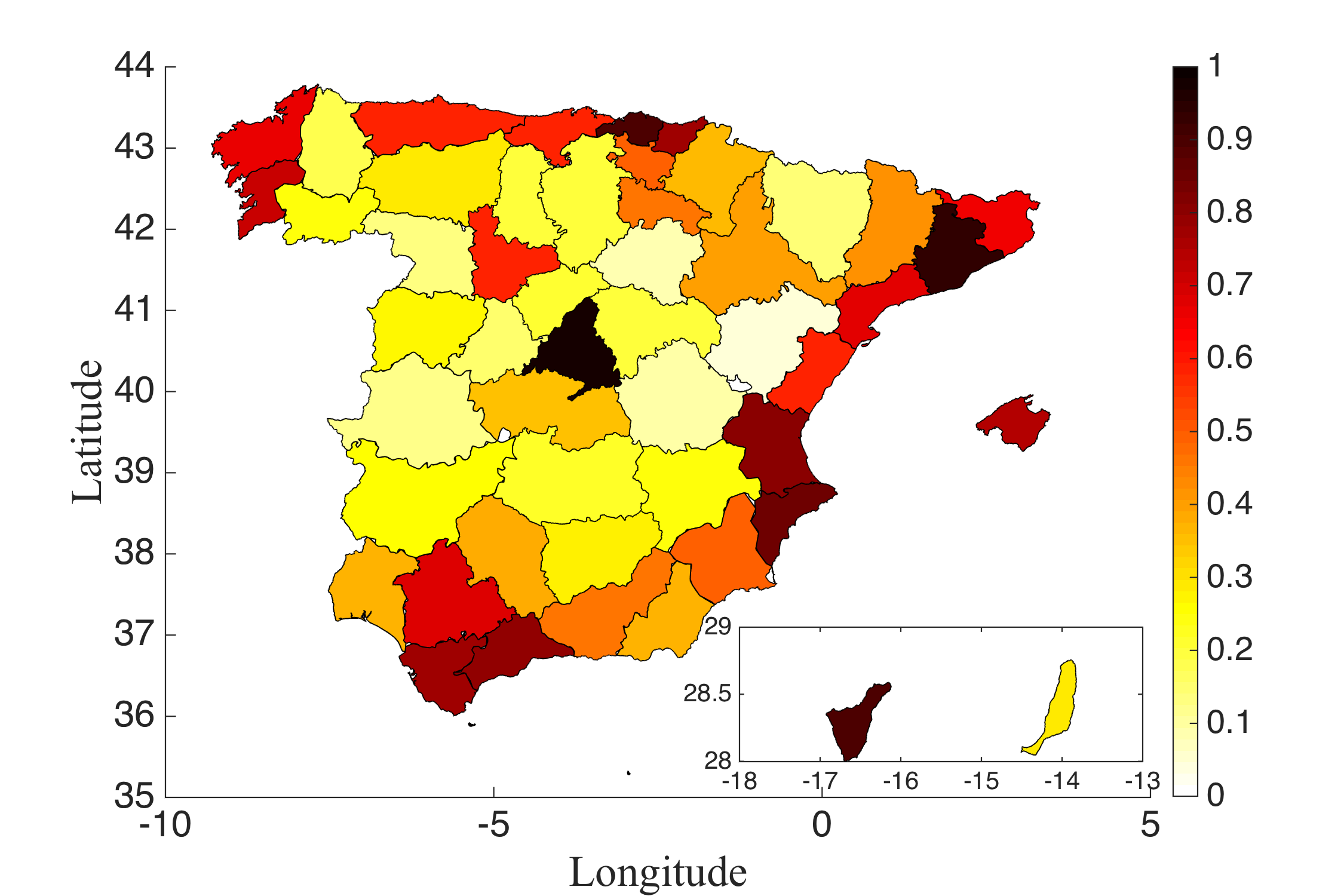}
  \caption{Spatial distribution of the density of total spending activity of domestic customers per $km^2$ of the province area on the normalized scale. The most economically dense Spanish provinces that have the highest spending density values are Madrid in the center and Barcelona -- the second province starting from right upper corner.}
 \label{fig:maps}
\end{figure}

We analyze the complete set of bank card transactions recorded by BBVA during 2011, all over Spain\footnote{Although the raw dataset is protected by a non-disclosure agreement and is not publicly available, certain aggregated data may be shared upon a request and for the purpose of findings validation.}. Spain has an area of 505,519 $km^2$ and counts $46,507,760$ inhabitants (2014). It is bordered to the northeast with France (which is separated from the chain of the Pyrenees) and Andorra, on the south by the Mediterranean Sea and Gibraltar (small possession of the United Kingdom) and, in Africa, with Morocco (through the autonomous cities of Ceuta and Melilla, its exclave). Spain is divided into $17$ autonomous communities (comunidades autónomas, singular: comunidad autónoma) which are further divided into $50$ provinces, plus $2$ autonomous cities: Ceuta and Melilla (officially designated as Plazas de Soberanía en el Norte de África). Gibraltar is claimed by Spain. Ceuta, Melilla and other small islands, which extend over $0.65$ $km^2$ and count $312$ inhabitants are the remains of the vast colonial empire that the country possessed. In total, Spain has $31.65$ $km^2$ of territory in North Africa, populated by $138,228$ inhabitants. We analyze the economic activity, during the 2011, over the $50$ provinces plus Ceuta and Melilla resulting in total of $52$ analyzed regions. \figurename~\ref{fig:maps} reports the density of total spending activity per $km^2$ in each province, which also serves as one of the indicators in the rest of the study.

Transactions that are in our dataset were performed by two groups of bank card users. The first one consists of BBVA direct customers, residents of Spain, who hold a debit or credit card issued by BBVA. In 2011, the total number of active customers was around 4.5 million, altogether they executed more than 178 million transactions in over 1.2 million points of sale, spending over 10 billion euros. The second group of bank card users includes over 8.6 million foreign customers of all other banks abroad coming from 175 countries, who made purchases through one of the approximately 300 thousand BBVA card terminals. In total, they executed additional 17 million transactions, spending over 1.5 billion euro.

Due to the sensitive nature of bank data, our dataset was anonymized by BBVA prior to sharing, in accordance to all local privacy protection laws and regulations. As a result, customers are identified by randomly generated IDs, connected with certain demographic characteristics and an indication of a residence location - at the level of zip code for direct customers of BBVA and country of residence for all others. Each transaction is denoted with its value, a time stamp, a retail location where it was performed, and a business category it belongs to. The business classification includes 76 categories, which were further aggregated into 12 meaningful major groups (e.g., purchases of food, fashion, home appliances or travel activities).

\section{Methodology}
\label{methodology}

The aim of this work is to examine whether bank card transactions can be used as good forecasters for macroeconomic quality of life indicators at \emph{mesoscale} level (i.e., provinces). As it has been already mentioned in Introduction, we use machine learning techniques to build a model whose inputs are microeconomic indicators extracted from BBVA dataset, while outputs present various indices denoting quality of life. In this way contribution of our paper is twofold: first we propose how to define and compute $35$ different microeconomic indicators based on individual bank card transactions and then we describe and validate the proposed approach for teaching a model based on them. 

\begin{table*}[b!]
 \caption{Economic indicators at province level extracted from the bank card transactions in Spain during the year 2011.}
 \vspace{-10pt}
\begin{center}
  \begin{tabular}{| c | p{15cm} |}
  \hline
  Indicator & Name \\ [1 pt] \hline 
    \hline
1&  Density of the spending activity within the area\\ [1 pt] \hline
2&  Density of the earnings within the area \\ [1 pt] \hline
3&  Average amount of a single transaction earned within the area\\ [1 pt] \hline
4&  Annual number of transactions per customer of the customers living in the area\\ [1 pt] \hline
5&  Amount of transaction performed by customers living within the area\\ [1 pt] \hline
6&	Percentage of area activity, received from out-of-province visitors \\ [1 pt] \hline
7&	Percentage of area activity, received from foreign visitors \\ [1 pt] \hline
8&	Area’s earning diversity\\ [1 pt] \hline
9&	Area’s spending diversity\\ [1 pt] \hline
10&	Area’s business density\\ [1 pt] \hline
11&	Average business size within an area\\ [1 pt] \hline
12&	Percentage of gas/parking/toll spending of area’s residents\\ [1 pt] \hline
13&	Percentage of taxi spending of area’s residents\\ [1 pt] \hline
14&	Percentage of public transportation spending of area’s residents\\ [1 pt] \hline
15&	Percentage of cafés/restaurants spending of area’s residents\\ [1 pt] \hline
16&	Percentage of fast food spending of area’s residents\\ [1 pt] \hline
17&	Percentage of food spending of area’s residents\\ [1 pt] \hline
18&	Percentage of recreation spending of area’s residents\\ [1 pt] \hline
19&	Percentage of fashion/beauty/jewelry spending of area’s residents\\ [1 pt] \hline
20&	Percentage of medical spending of area’s residents\\ [1 pt] \hline
21&	Percentage of cultural spending of area’s residents\\ [1 pt] \hline
22&	Percentage of travel spending of area’s residents\\ [1 pt] \hline
23& Percentage of area’s residents nighttime spending\\ [1 pt] \hline    
24&	Percentage of area’s residents weekend spending \\ [1 pt] \hline
25&	Percentage of area’s residents nighttime money spending\\ [1 pt] \hline  
26&	Percentage of area’s residents weekend money spending\\ [1 pt] \hline
27&	Percentage of area’s nighttime earnings\\ [1 pt] \hline 
28&	Percentage of area’s residents weekend earnings\\ [1 pt] \hline
29&	Percentage of area’s nighttime business transactions\\ [1 pt] \hline  
30& Percentage of area’s residents weekend transactions\\ [1 pt] \hline
31& Percentage of area residents' activity performed outside the province\\ [1 pt] \hline 
32& Percentage of out area residents' activity performed inside the province\\ [1 pt] \hline
33& Percentage of money spent by area residents outside their province\\ [1 pt] \hline
34& Percentage of money spent by out of area residents inside a province\\ [1 pt] \hline
35&	Area’s residents spending in expensive locations.\\ [1 pt] \hline
  \end{tabular}
\end{center}
\label{tab:indicators}
\end{table*}
\subsection{Microeconomic indicators}
\label{ind}

From BBVA dataset of individual spending behavior for the period of one year, we extract $35$ different microeconomics indicators that explain economic behaviors from both customer and business sides (see Table~\ref{tab:indicators}). Before calculating the aforementioned parameters, BBVA dataset had to be pre-processed in order to compensate for potential bias introduced by the spatial inhomogeneity of BBVA market share. The first concern was: what is BBVA penetration in the whole banking market for the given area (i.e., what is the ratio of BBVA customers and economically active population)? Therefore, in order to estimate the total domestic customer spending volume, customers' activity was normalized by the bank's market share corresponding to their residence location and grouped at the level of provinces. Another type of bias is related to unequal distribution of foreign customers performing transactions in BBVA point of sale terminals in different locations across the whole Spain. In this case the normalization procedure relied on BBVA business market share defined, for the purpose of this study, as a ratio of bank card transactions executed by domestic customers in BBVA terminals and their transactions in all other terminals located in the considered area. The appropriate normalization allows estimation of the total spending volume of foreign customers visiting a particular location.

The indicators at province scale showed in Table~\ref{tab:indicators} can be split in two macro-categories: (i) customer and (ii) business (i.e., merchant) side. For instance the first eleven indicators refer to the economic activity inside each province. Indicator 1 has been computed by evaluating the average density of number of transactions made within 1 $km^2$ of the province area, while Indicator 2 refers to the average density of amount of money spent, and Indicator 3 denotes the ratio between total amount and number of transactions made by all customers within the considered area. 

Indicators 4, 5 and 6 are more focused on the customer side. Indicator 4 evaluates the average number of transactions per customer, i.e., the ratio between the total number of transactions made by residents of the area and the number of active residents in terms of transaction activity. Indicator 5 computes the fraction between the total amount and the number of transactions made by residents of the considered area everywhere in the country, while Indicator 6 evaluates the percentage of the number of transactions made within the area by its domestic out-of-province visitors. Moreover, we also evaluated the effect of the foreign activity by considering the percentage of the number of transactions made within the area by its foreign visitors. 

In order to also include the effect of the structure of activity by its type, we consider something that what we call -- earning and spending \emph{diversity}. In that sense, Indicators 8 and 9 represent the spending categorical diversity, specifically the number of top business categories (of 76) enough to cover 80\% of the total activity of area residents or activity within the area, respectively. Additionally, Indicator 10 reflects the number of active businesses within the area per $km^2$. For Indicator 11 we compute the average earnings of an active business within the area (i.e., the total earned amount divided by the number of active businesses). 

Indicators 12 to 22 correspond to the specific properties of the structure of spending activity within the area taking into account spending in different business categories, such as food, taxi, public transportation, etc. Finally, we evaluate the effect of the temporal activity by distinguishing nighttime and weekend temporal windows. For the purpose of defining Indicators 23 to 30 we assume that nighttime activity happens between 10 PM and 6 AM, while weekend activity counts for transactions made on Saturdays and Sundays. Indicators 31 to 34 reflect the customer activity inside or outside their provinces. The last indicator computes the percentage of the total transaction of residents made in the "expensive" businesses, i.e., those which average transaction amount is above average for the corresponding business category.

\subsection{Macroeconomic indices}

As mentioned in Introduction, a huge number of indicators can be used to characterize quality of life for whole countries and their citizens. In this work we decided to focus on six socioeconomic indices for the year 2011 that are available for Spanish province level and that are included in official Spanish statistic reports from Instituto Nacional de Estadística\footnote{http://www.ine.es} and Eurostat\footnote{http://ec.europa.eu/eurostat} web pages: GDP, housing price level, unemployment rate, crime rate, percentage of higher education, and life expectancy.


We choose GDP as it is widely used as a benchmark of successful public policy initiatives and as the primary objective of the lending decisions of major global economic institutions. The advantage of GDP is that it measures the aggregate economic activity within a country, but the downside is that economic activity generated for whatever purpose (e.g., building prisons or schools, spending more on health care, whether or not it is medically beneficial) raises GDP in the same way. 

In addition to economic indices, we also use social ones that are compiled by the Statistics Division, Department of Economic and Social Affairs of the United Nations Secretariat\footnote{http://www.un.org/en/development/desa/index.html} using many different national and international sources. Namely, the indices presented in this paper consist mainly of the minimum list that has been proposed for follow-up and monitoring implementation of major United Nations conferences on children, population and development, social development and women. 

This minimum list is contained in the Report of the Expert Group on the Statistical Implications of Recent Major United Nations Conferences (E/CN.3/AC.1/1996/R.4). Technical background on the development of social indices is available in two United Nations publications: \textit{Handbook on Social Indicators (United Nations publication, Series F, No. 49, 1989)} and \textit{Towards a System of Social and Demographic Statistics (United Nations publication, Series F, No. 18, 1975)}\footnote{http://unstats.un.org/unsd/demographic/products/socind}. All aforementioned indices are provided for the following areas: population, health, housing, education and work.

\subsection{The model}

\begin{figure}[b!]
  \centering
      \includegraphics[width=1\columnwidth]{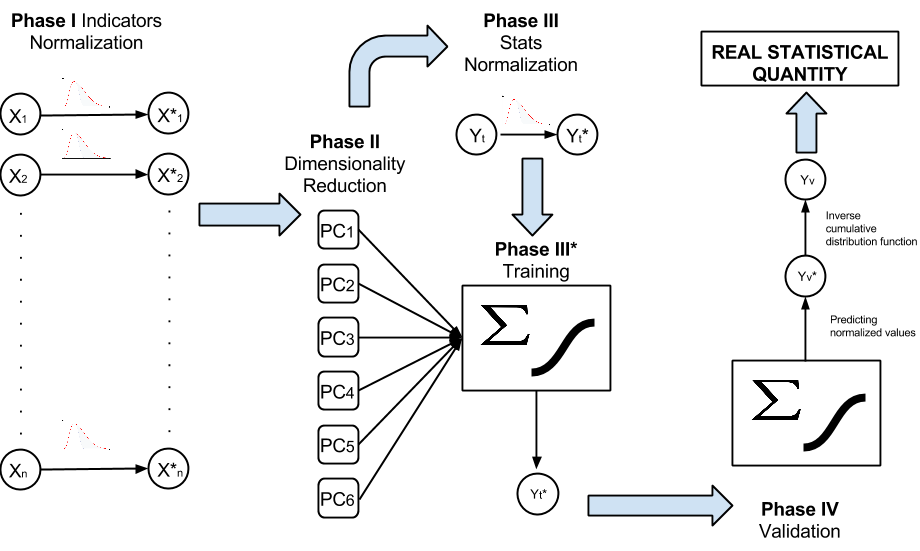}
  \caption{Schematic representation of the model used in this research. }
\label{fig:ml}
\end{figure}

The first step in building our model is to normalize all micro- and macroeconomic parameters to be between $0$ and $1$ by fitting an appropriate distribution (normal or lognormal whichever fits better) as shown in~\figurename~\ref{fig:ml}. In doing so, we transformed the data using cumulative distribution function of fitted distribution (i.e., replace the original data with the corresponding quantile values). This is similar to the quantile normalization introduced in \cite{Bolstad2003} but instead of using the certain empirical distribution, in this paper we use the actual best-fit distribution function. For each indicator we evaluate which distribution fits better following maximum-likelihood estimation:

\begin{equation}
l = \sum_i ^N ln f(x_i)
\end{equation}

\noindent
where $f$ denotes the considered partial distribution function. We choose the distribution and its parameters maximizing $l$.

\begin{figure*}[t!]
\centering
        \begin{subfigure}[b]{0.5\textwidth}
                \includegraphics[width=\linewidth]{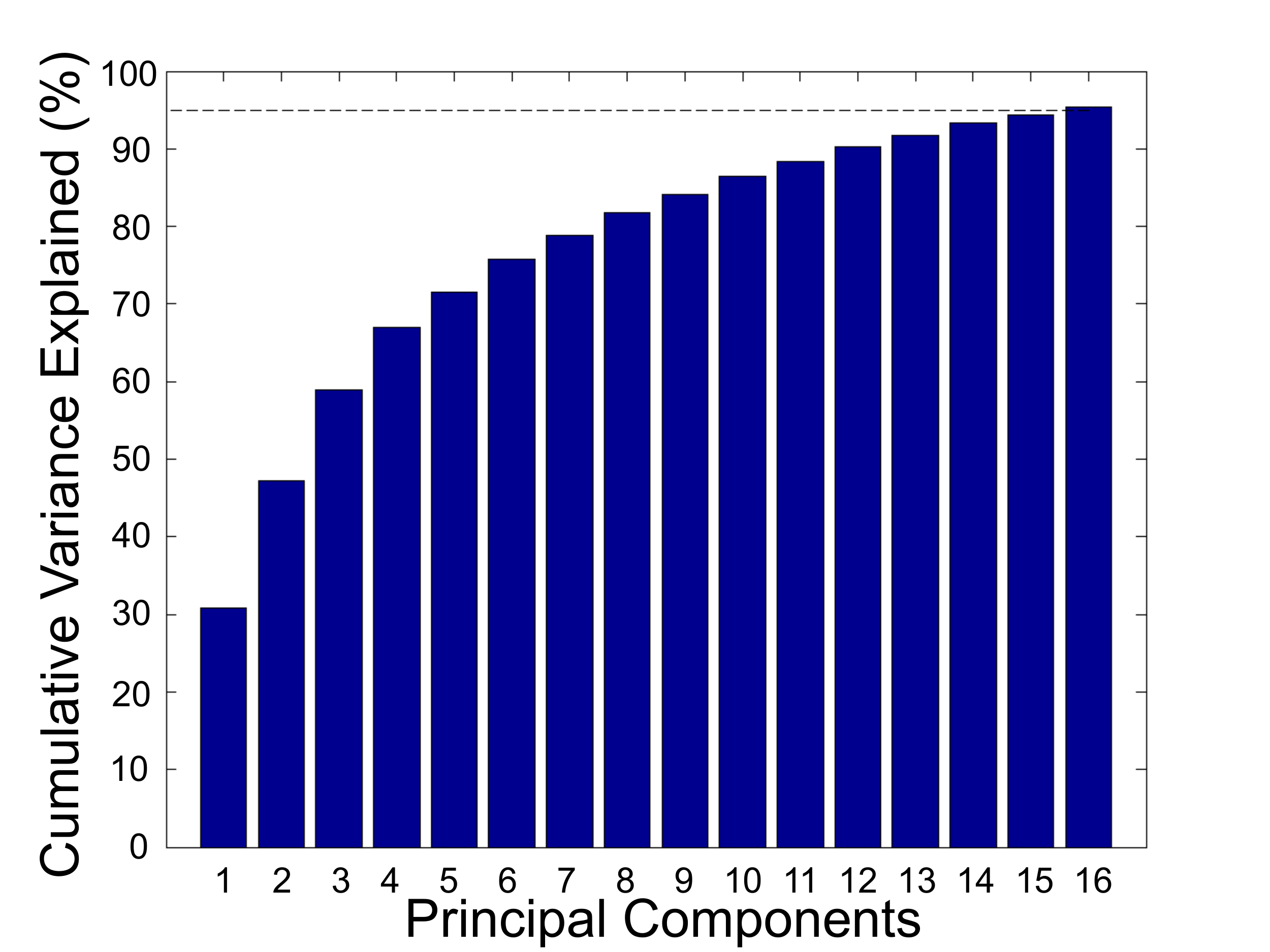}
                \label{fig:pca0}
        \end{subfigure}%
        \begin{subfigure}[b]{0.5\textwidth}
                \includegraphics[width=\linewidth]{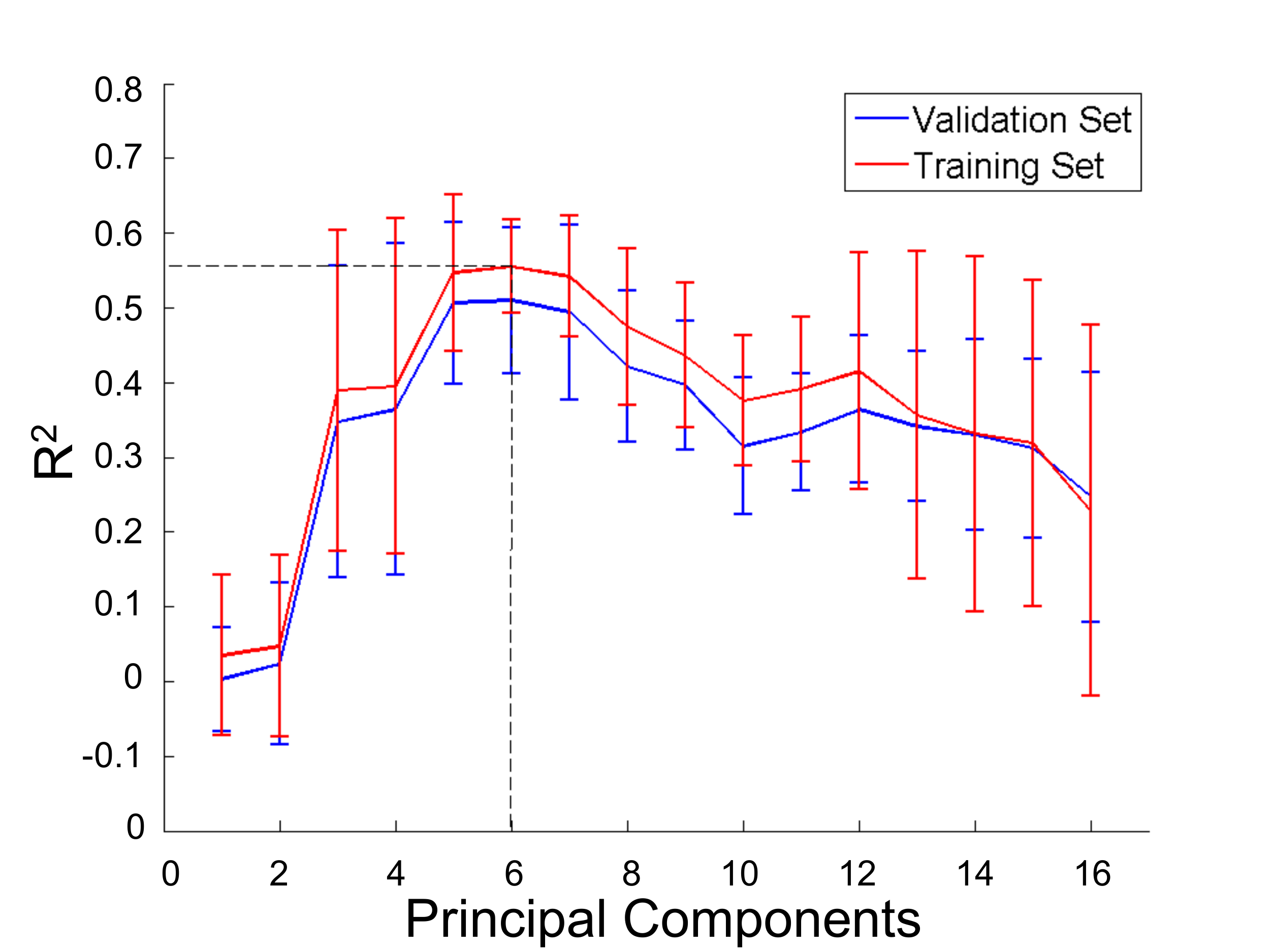}
                \label{fig:pca1}
        \end{subfigure}%
        \caption{(a) We show the first $16$ of the total $35$ principle components explain $95\%$ of the total data variance. (b) However, using only $6$ principal components allows to reach the best results in terms of model fitting/cross validation.}
        \label{fig:pca}
\end{figure*}

Since many of the indicators are strongly correlated with each other, the next step is to perform dimensionality reduction using standard Principle Component Analysis (PCA). PCA can be used to compress the information from a large number of variables to a smaller dataset while minimizing the information lost during this process \cite{jolliffe2002principal}. Setting a threshold for the total percentage of information that should be covered (we use $95\%$), we get a reasonable selection of the top independent components. Those components "saved" most of data that all individual indicators together provided before the reduction process as shown in \figurename~\ref{fig:pca}a. The result reported here is for all 52 provinces together, while in our further analysis, which is presented in Section \ref{res}, we will be considering different training sets for our model, all of them being subsets of the entire one. However, the results for those subsets do not differ substantially from the results presented here for the whole set.

Selected principle components are then used as a feature space for teaching our model to predict quality of life parameters at the province level in Spain. As mentioned before this model can be further applied for predicting quality of life parameters on much more fine-grained spatial scales (e.g., cities, districts and smaller neighborhoods) for which consistent official statistics does not exist. After principle components were selected, we analyze their individual correlations with macroeconomic statistical parameters to be predicted, in order to see if in theory we can get a decent predictive power using the feature space we built. The correlation is given by a matrix $R$ of correlation coefficients calculated from our input microeconomics and output macroeconomic parameters. The matrix $R$ is related to the covariance matrix C as:

\begin{equation}
R(i,j) = \frac{C(i,j)}{\sqrt{C(i,i)C(j,j)}}.
\end{equation}

\begin{figure*}[t!]
\centering
        \begin{subfigure}[b]{0.33\textwidth}
                \includegraphics[width=\linewidth]{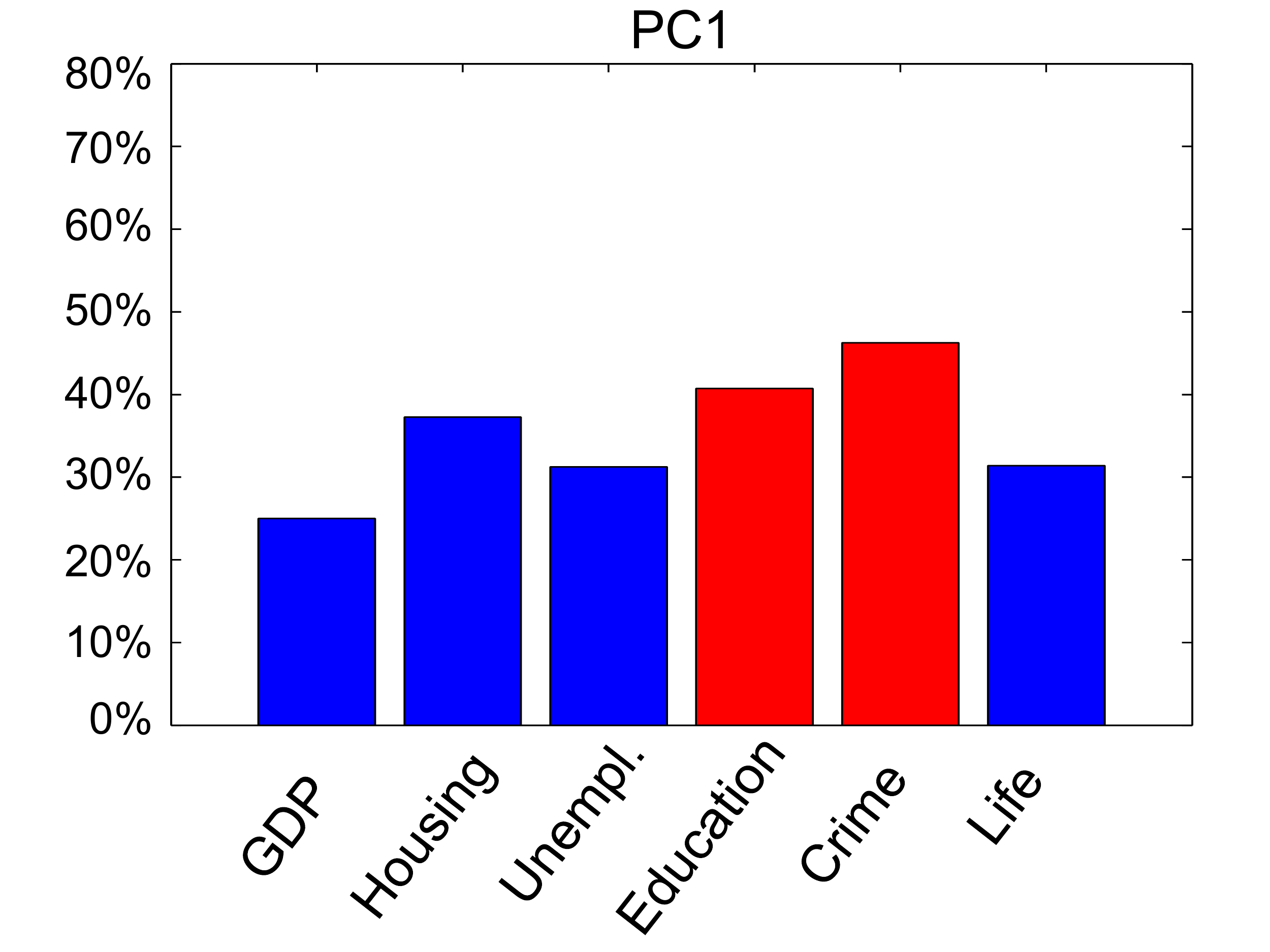}
                \label{fig:pc1}
        \end{subfigure}%
        \begin{subfigure}[b]{0.33\textwidth}
                \includegraphics[width=\linewidth]{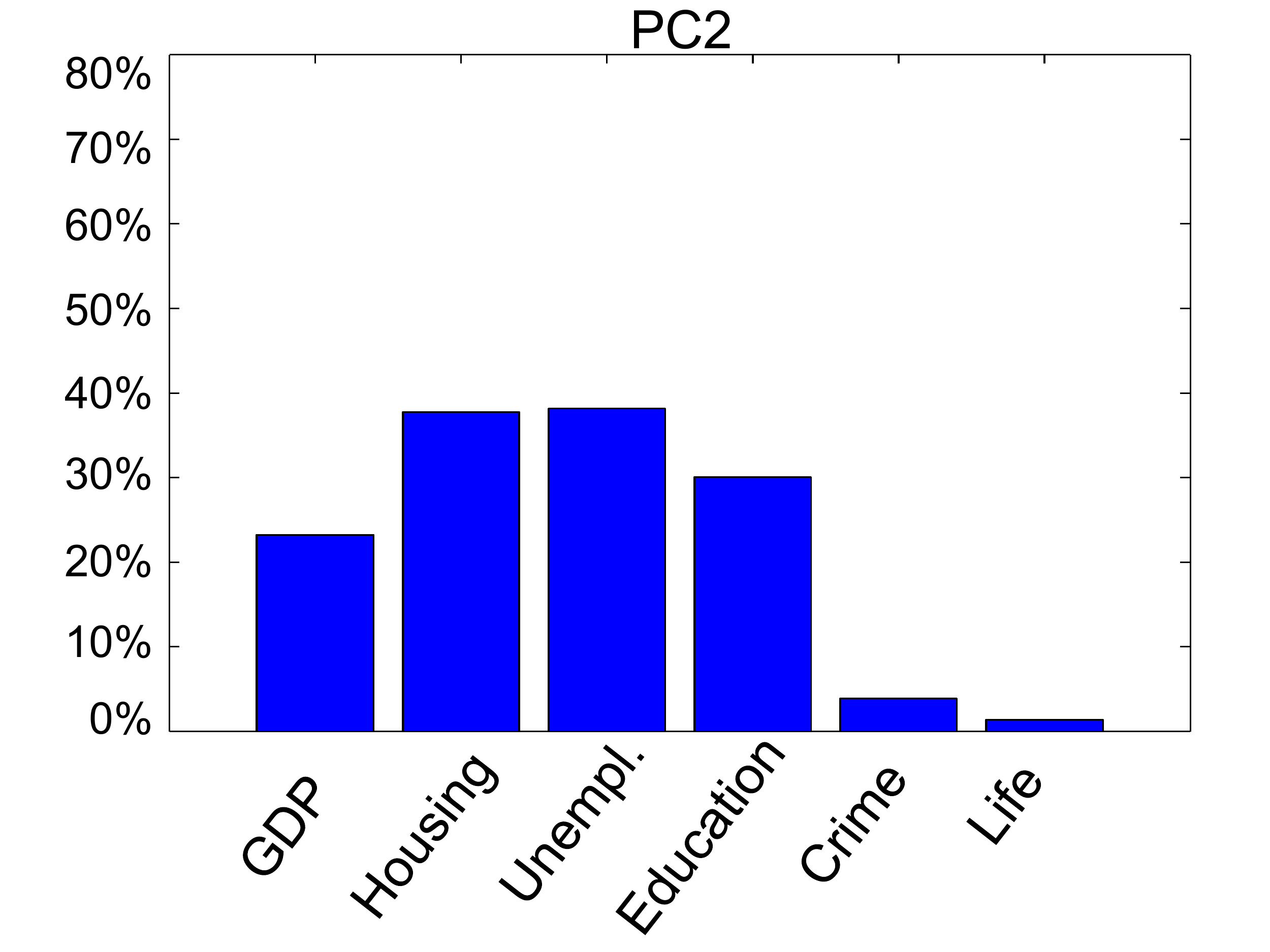}
                \label{fig:pc2}
        \end{subfigure}%
        \begin{subfigure}[b]{0.33\textwidth}
                \includegraphics[width=\linewidth]{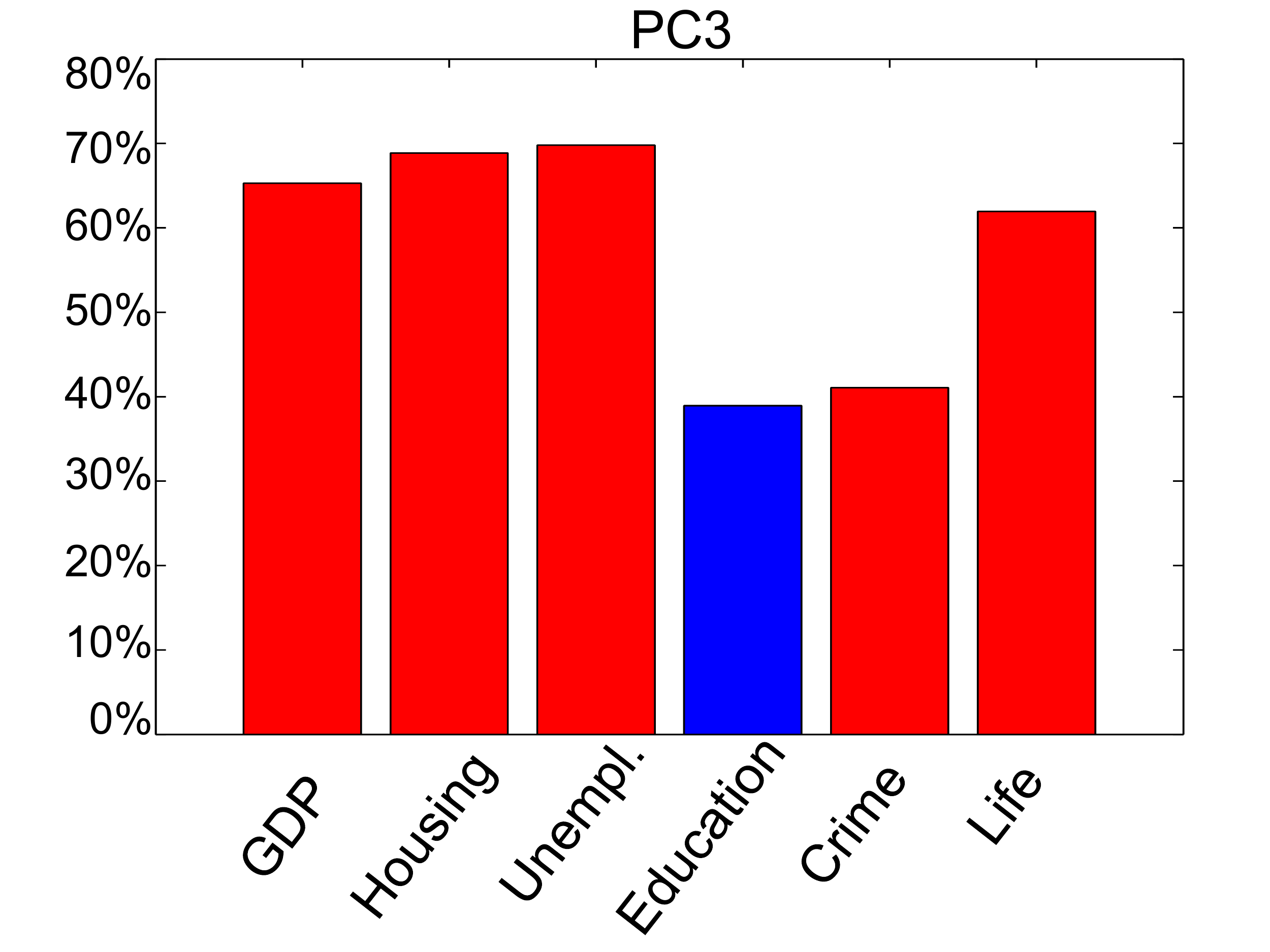}
                \label{fig:pc3}
        \end{subfigure}%
        \\
        \vspace{-10pt}
        \begin{subfigure}[b]{0.33\textwidth}
                \includegraphics[width=\linewidth]{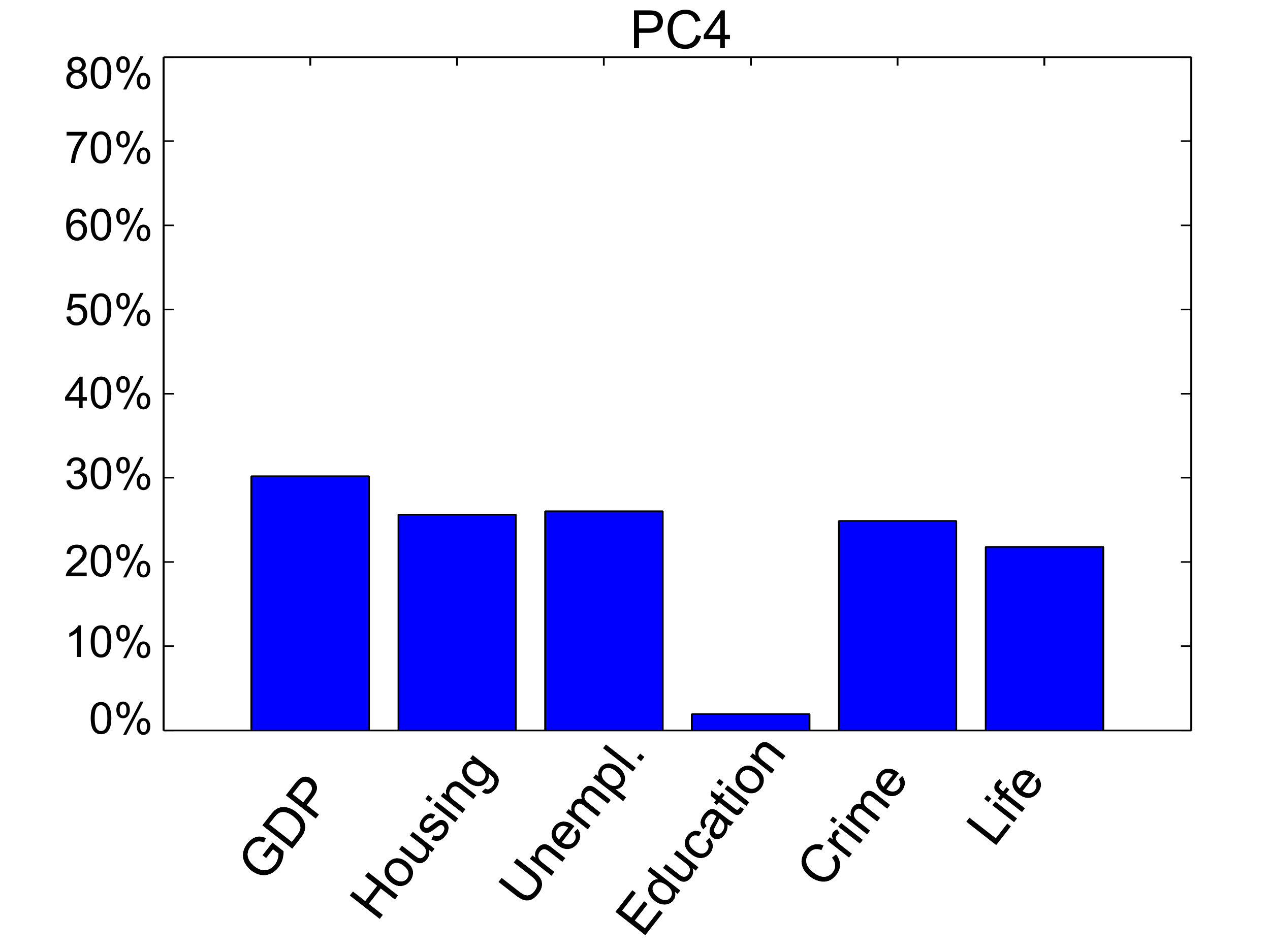}
                \label{fig:pc4}
        \end{subfigure}%
        \begin{subfigure}[b]{0.33\textwidth}
                \includegraphics[width=\linewidth]{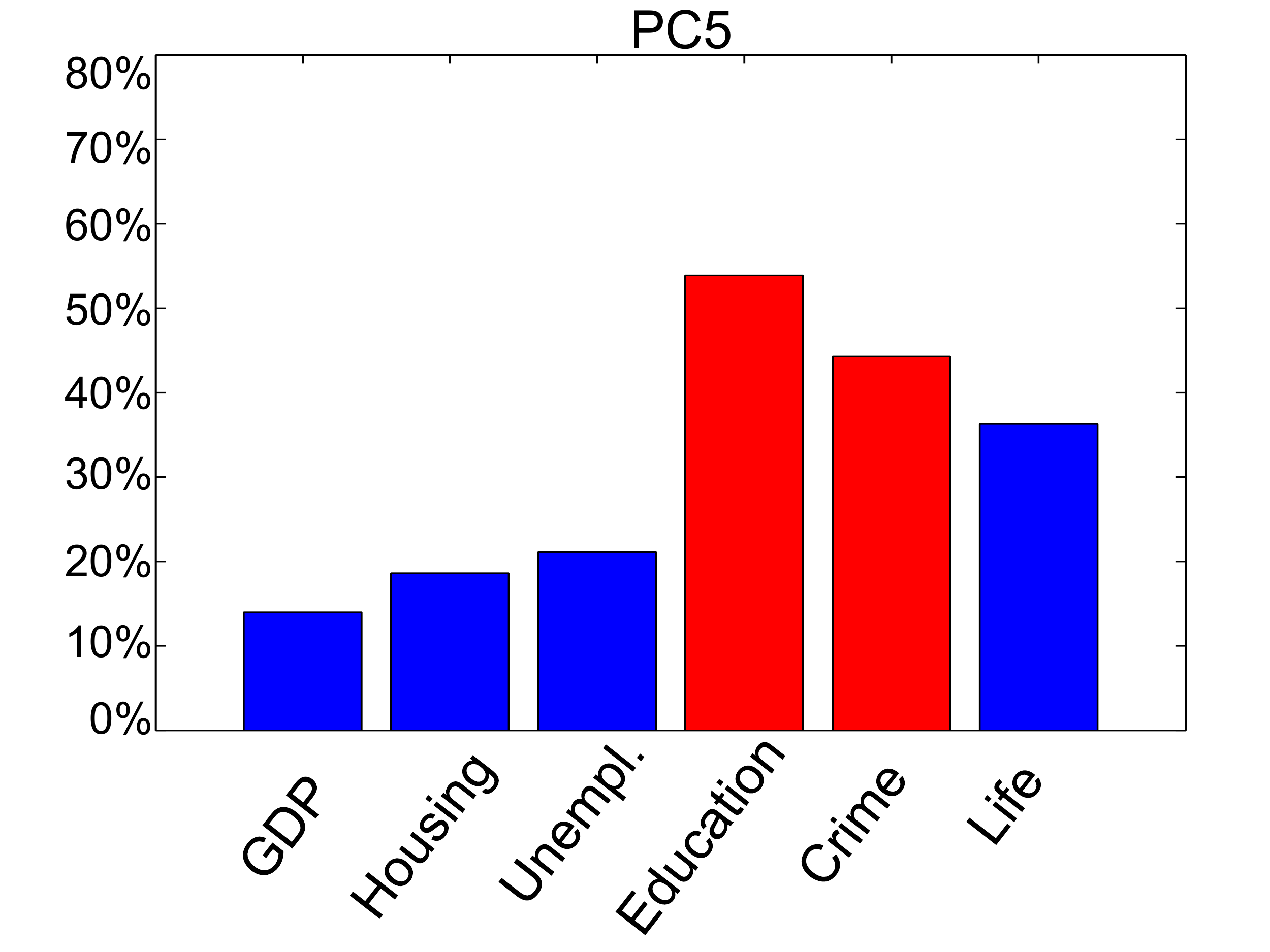}
                \label{fig:pc5}
        \end{subfigure}%
        \begin{subfigure}[b]{0.33\textwidth}
                \includegraphics[width=\linewidth]{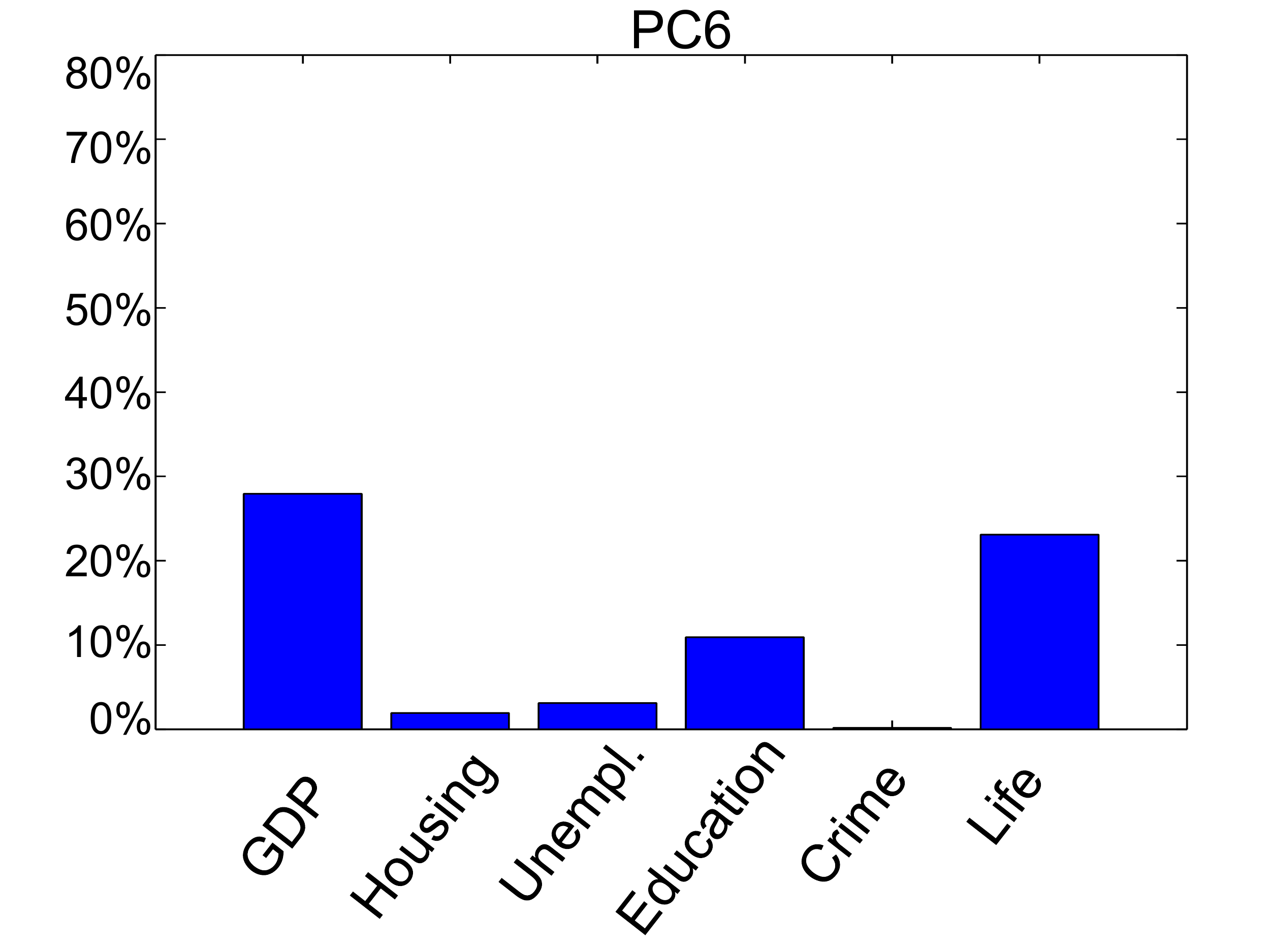}
                \label{fig:pc6}
        \end{subfigure}%
        \vspace{-10pt} 
        \caption{Absolute values of individual correlations between $6$ principal components and $6$ statistical parameters. Red bars indicate the highest correlations (i.e., above $40\%$).}
        \label{fig:correlation}
\end{figure*}

The correlations between $16$ principal components, extracted from $35$ indicators, and $6$ socioeconomic statistical indices for all 52 provinces are shown in Table~\ref{tab:correlation}. Strongest positive or negative correlations are also highlighted by red bars in \figurename~\ref{fig:correlation}. The first principal component is mostly correlated with the percentage of higher education and crime rate and slightly weaker --- with housing prices, the third --- with literally all the quantities, the fifth --- mostly with education rate, while second and fourth show rather weaker correlations with our six statistical indices. Moreover, other principle components starting from the sixth one already show pretty insignificant correlations with the statistical parameters considered in this paper, showing that not all of them have the same strong impact on the model performance and that we should decide which ones to use.

\begin{table}[b!]
 \caption{Individual correlations between $16$ microeconomic (principal components) and $6$ macroeconomic parameters.}
 \small{
\begin{center}
  \begin{tabular}{| r | r | r | r | r | r | r |}
  \hline
  PC & GDP & Housing & Unempl. & Educ. & Crime & Life\\ \hline
    \hline
  1 & 24.99 & 37.27 & 31.24 & 40.73 & 46.25 & -31.40 \\
  2 & 23.26 & -37.72 & -38.15 & 30.04 & 3.89 & 1.37 \\
  3 & 65.29 & -68.87 & -69.79 & 38.95 & -41.07 & 61.94 \\
  4 & -30.19 & 25.61 & 26.01 & -1.94 & 24.87 & -21.79 \\
  5 & -13.99 & 18.60 & 21.11 & -53.89 & 44.27 & -36.29 \\
  6 & -27.93 & 1.94 & 3.12 & -10.93 & -0.18 & -23.09 \\
  7 & 6.00 & -1.93 & -1.93 & -7.71 & -7.73 & 23.55 \\
  8 & 24.17 & -23.91 & -24.82 & 21.16 & -6.49 & 1.45 \\
  9& 10.69 & -5.13 & -7.27 & 28.73 & 1.09 & -2.14 \\
  10 & -27.75 & 22.16 & 23.77 & -22.74 & 1.57 & -19.20 \\
  11 & 5.20 & -12.39 & -12.23 & 4.10 & -17.81 & -4.20 \\
  12 & -0.80 & 1.26 & 2.25 & -1.31 & -20.95 & 5.03 \\
  13 & -9.62 & -8.38 & -6.06 & -5.48 & -7.19 & 26.63 \\
  14 & -5.08 & -11.46 & -12.26 & 10.13 & -17.63 & -2.30 \\
  15 & 6.33 & -12.84 & -12.16 & 3.30 & 8.30 & -1.16 \\
  16 & -20.87 & 21.39 & 21.28 & 1.48 & 4.19 & -21.08 \\ \hline
  \end{tabular}
\end{center}
}
\label{tab:correlation}
\end{table}

After doing PCA, the next step in building of our model process is to teach the model using the selected feature space that explains the statistical quantities at the considered spatial scale. We experimented with a standard Generalized Linear Model (GLM) using a logistic regression algorithm, as logistic function is typically applied in case of the binary or normalized values. GLM is a flexible generalization of ordinary linear regression that allows for response variables that have error distribution models other than a normal distribution. This model generalizes linear regression by allowing the linear model to be related to the response variable via a link function and by allowing the magnitude of the variance of each measurement to be a function of its predicted value~\cite{nelder1972generalized}. 

A schematic representation of the GLM used in this research has already been shown in Figure~\ref{fig:ml}. In the first step of doing GLM, all indicators are normalized between 0 and 1 applying the cumulative distribution function of the best-fit distribution they follow. Moreover, our model predicts values on the normalized scale resulting in that for each training session we also normalize output variables for each province (i.e., official statistical indexes). In the second step of this process, we compute a dimensionality reduction following PCA. In the third step we teach GLM model on the available data sample used as a training set. Note here, that as described earlier, different subsets of the data will be used further for this purpose. Then, for every sample point (i.e., province) certain number of principal components are used as input variables of the algorithm, while the output is the normalized statistical value for that province. The linear model is represented by the \emph{summation} symbol and the \emph{logit} function is represented by the \emph{sigmoid} curve symbol. Finally, we validate the model using the validation set (i.e., remaining provinces) and apply the corresponding inverse cumulative distribution function to evaluate the predicted quantities on the original scale from the predicted normalized values.

In standard \emph{linear models} the expected value of the response variable $\mathbf{Y} \in \mathbf{R}^m$ is supposed to linearly depend on its coefficient, $\mathbf{\beta} \in \mathbf{R}^n$ acting upon the set of $n$ predictor variables $\mathbf{X} \in \mathbf{R}^{n \times m}$:

\begin{equation}
E(\mathbf{Y}) = (\mathbf{\beta}^T\mathbf{X})^T,
\end{equation}

\noindent
while the standard GLM model, first developed by Nelder and Weddeburn \cite{nelder1972generalized}, takes a more general form:

\begin{equation}
E(\mathbf{Y}) = g^{-1}\left((\mathbf{\beta}^T\mathbf{X})^T \right)
\end{equation}

\noindent
with the response variable, $\mathbf{Y} | \mathbf{\beta}^T\mathbf{X}$, belonging to a specified distribution from a single parameter exponential family and $g^{-1}(\cdot)$ providing an appropriate transformation from the linear predictor, $\mathbf{Y} | \mathbf{\beta}^T\mathbf{X}$, to the conditional mean, $\mu$. The inverse of the mean function, $g^{-1}(\cdot)$, is known as the \emph{link function} $g^{-1}$. In this paper we use the link function \emph{logit} \cite{nelder1972generalized}:

\begin{equation}
g(p)=log\frac{p}{1-p}.
\end{equation}
\begin{figure*}[t!]
\centering
        \begin{subfigure}[b]{0.47\textwidth}
                \includegraphics[width=\linewidth]{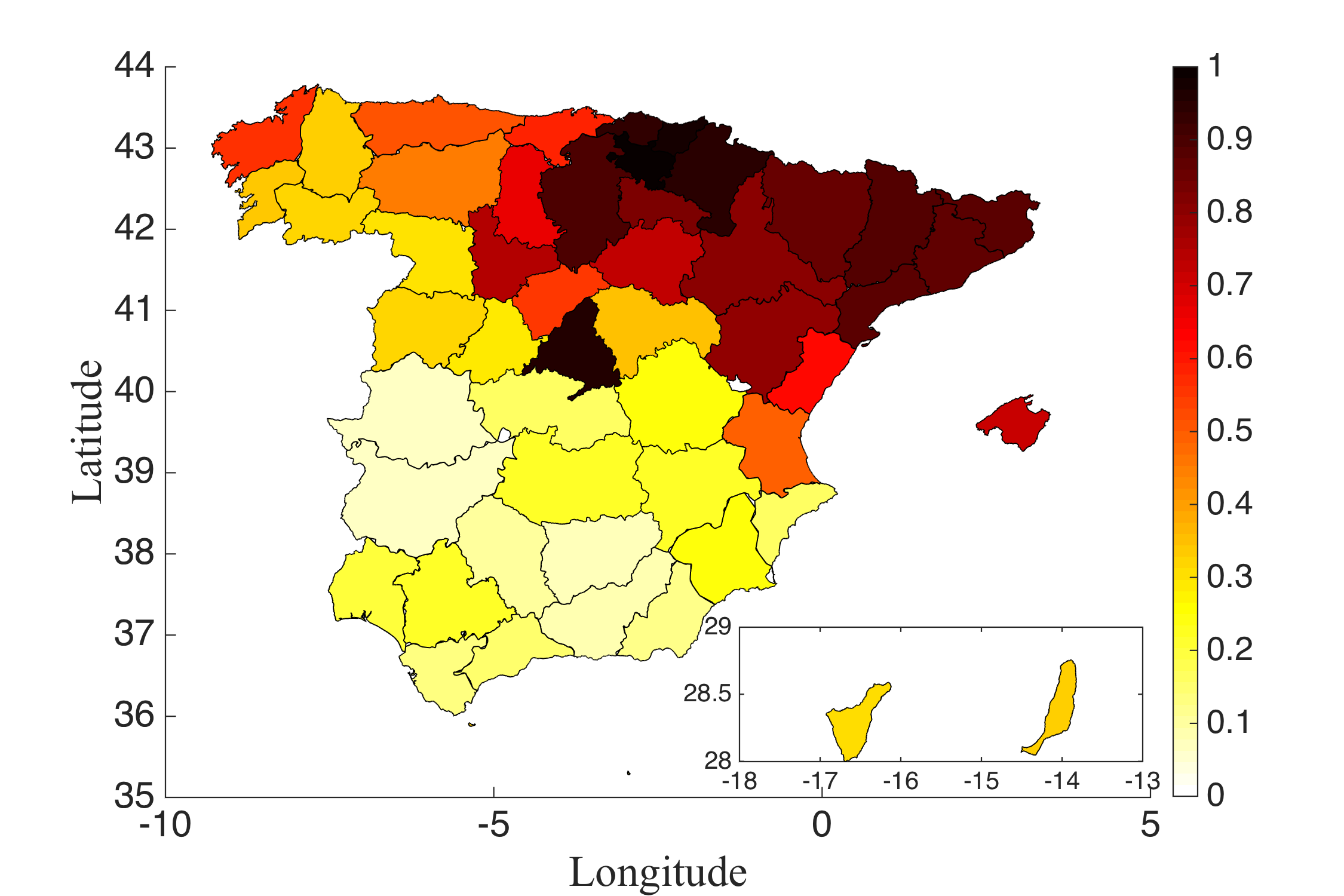}
                \label{fig:pca0}
        \end{subfigure}%
        \begin{subfigure}[b]{0.47\textwidth}
                \includegraphics[width=\linewidth]{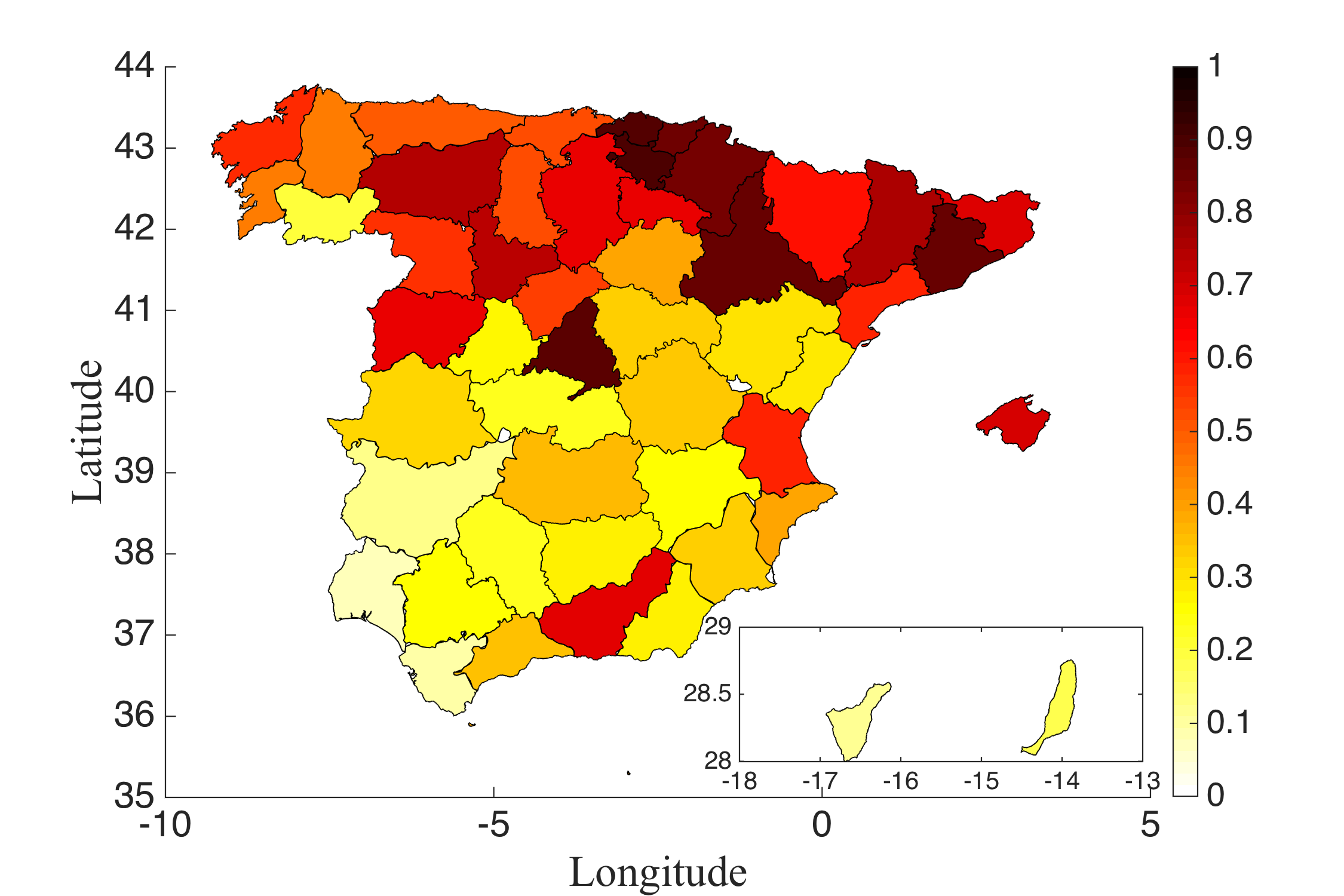}
                \label{fig:pca1}
        \end{subfigure}%
        \caption{Spatial distribution of the normalized GDP values (a) and model prediction (b). Here the model has been trained on all $52$ provinces.}
        \label{fig:gdp_map}
\end{figure*}

\begin{figure*}[t!]
\centering
        \begin{subfigure}[b]{0.4\textwidth}
                \includegraphics[width=\linewidth]{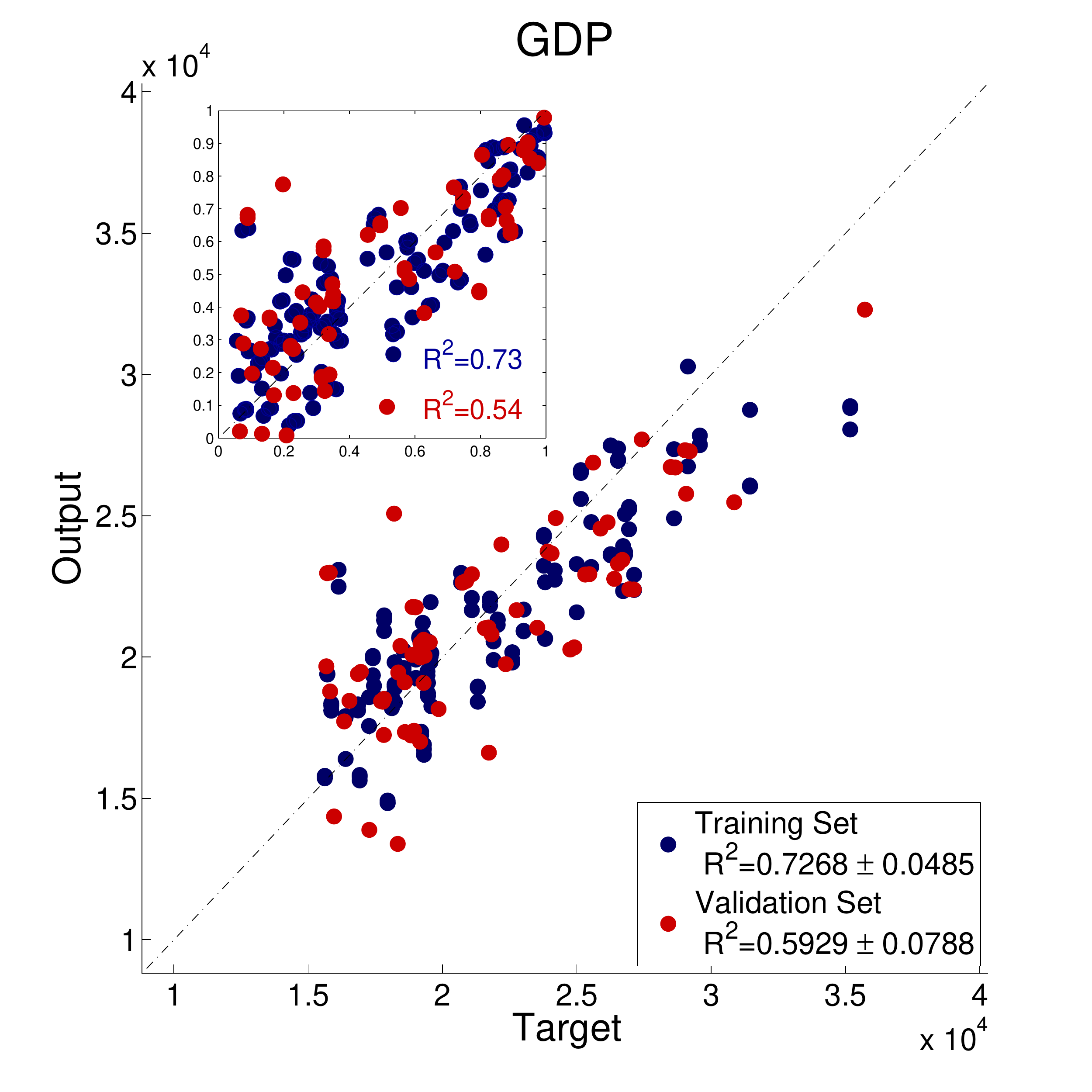}
                \label{fig:gdp}
        \end{subfigure}%
        \begin{subfigure}[b]{0.4\textwidth}
                \includegraphics[width=\linewidth]{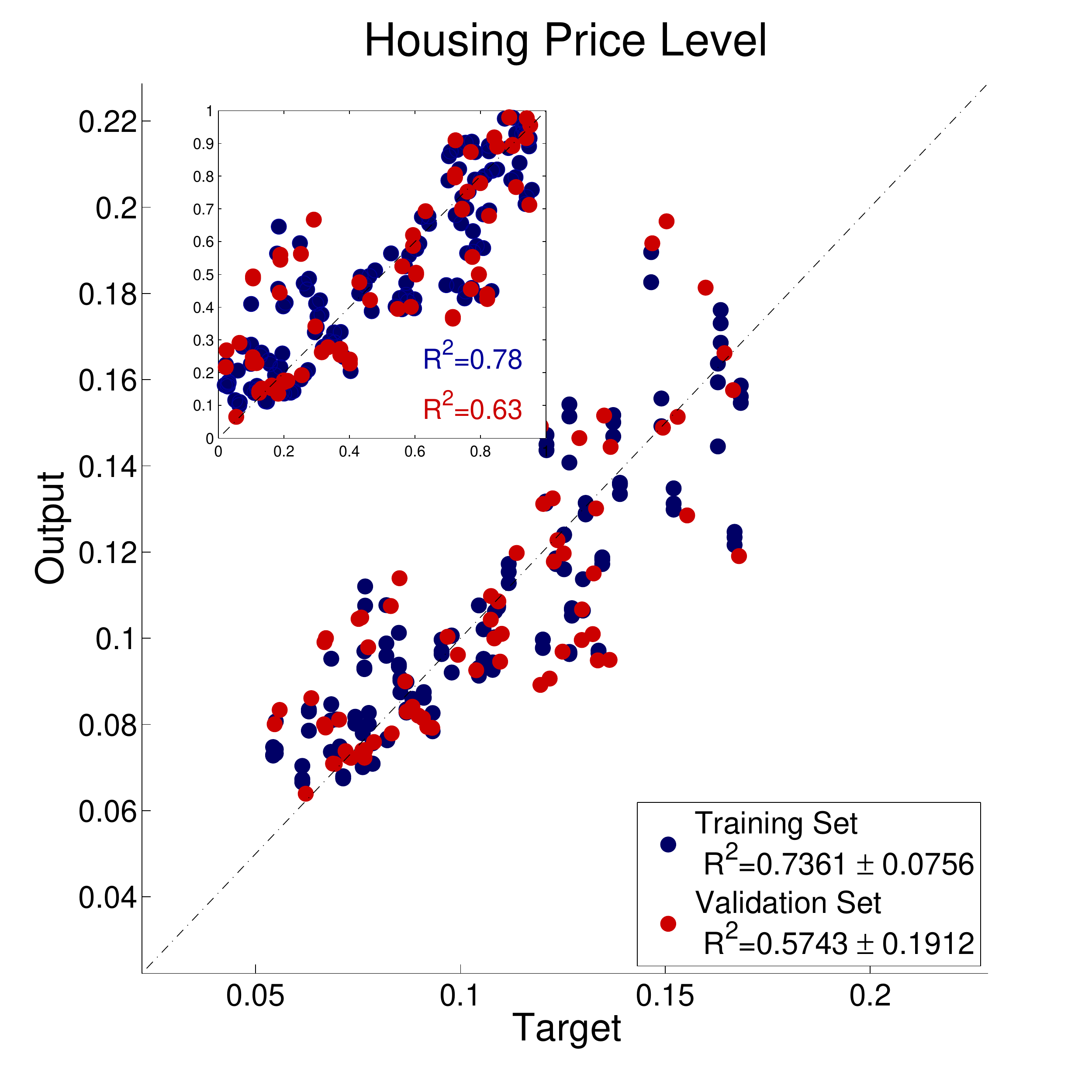}
                \label{fig:housing}
        \end{subfigure}%
        \\
        \begin{subfigure}[b]{0.4\textwidth}
                \includegraphics[width=\linewidth]{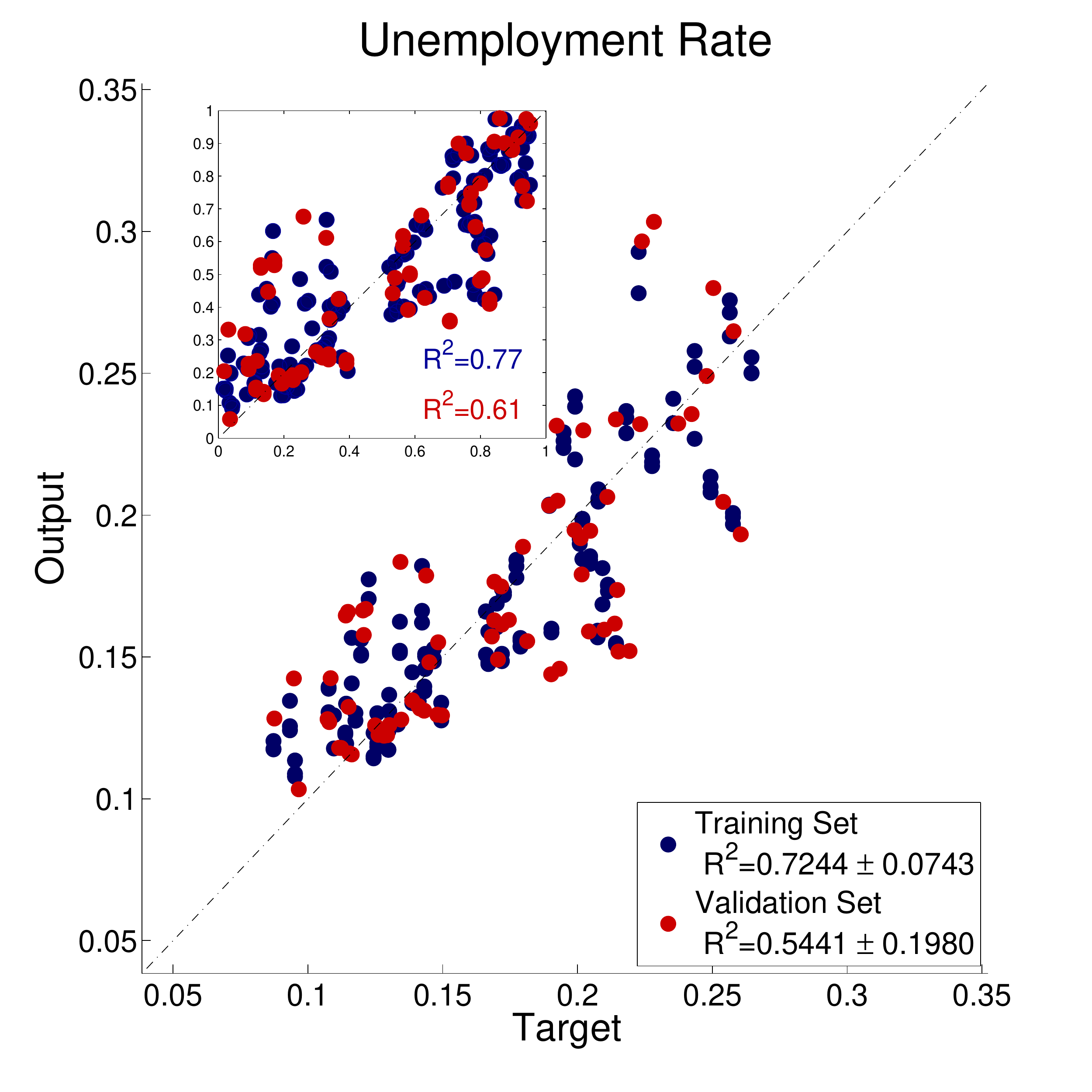}
                \label{fig:unemp}
        \end{subfigure}%
        \begin{subfigure}[b]{0.4\textwidth}
                \includegraphics[width=\linewidth]{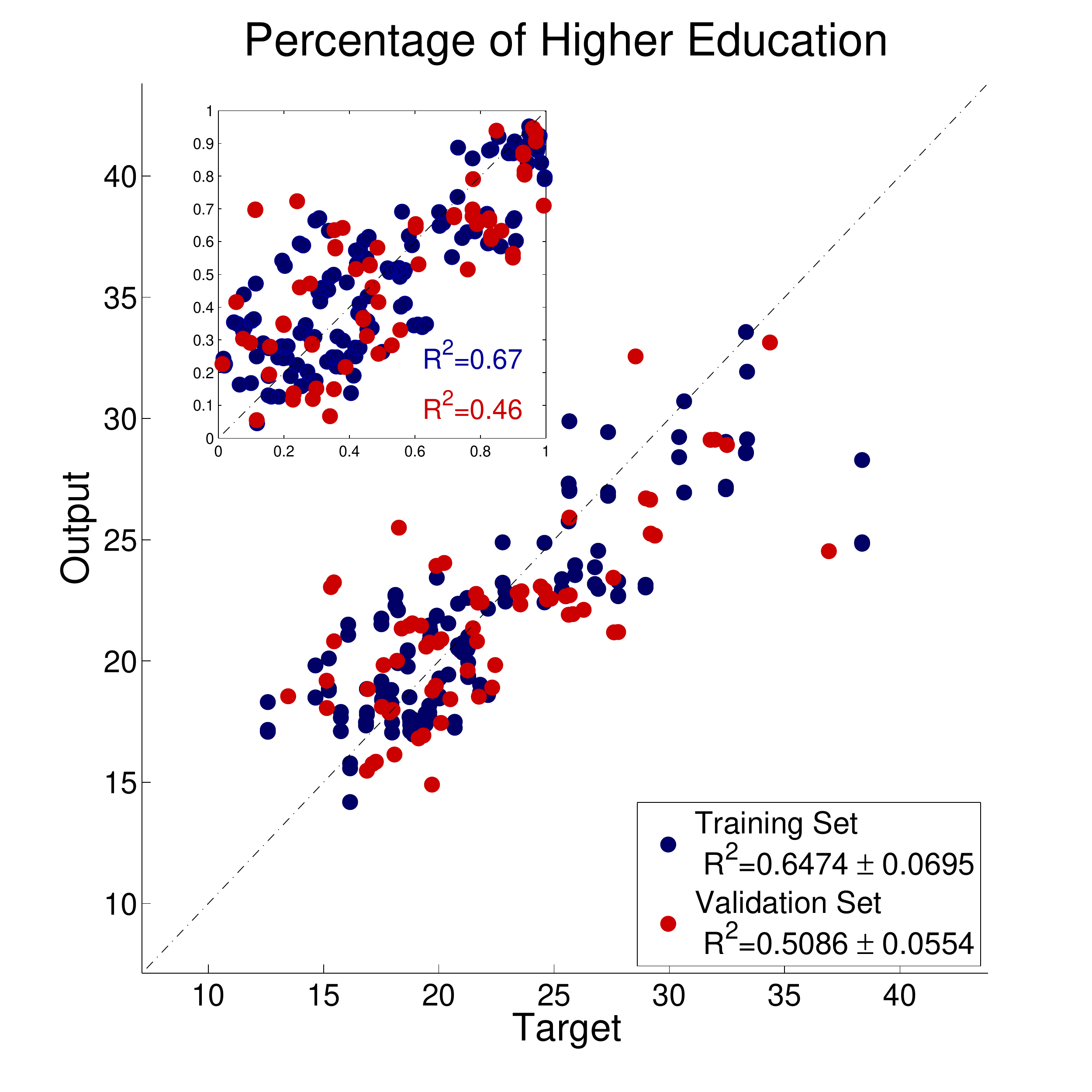}
                \label{fig:edu}
        \end{subfigure}%
        \\
        \begin{subfigure}[b]{0.4\textwidth}
                \includegraphics[width=\linewidth]{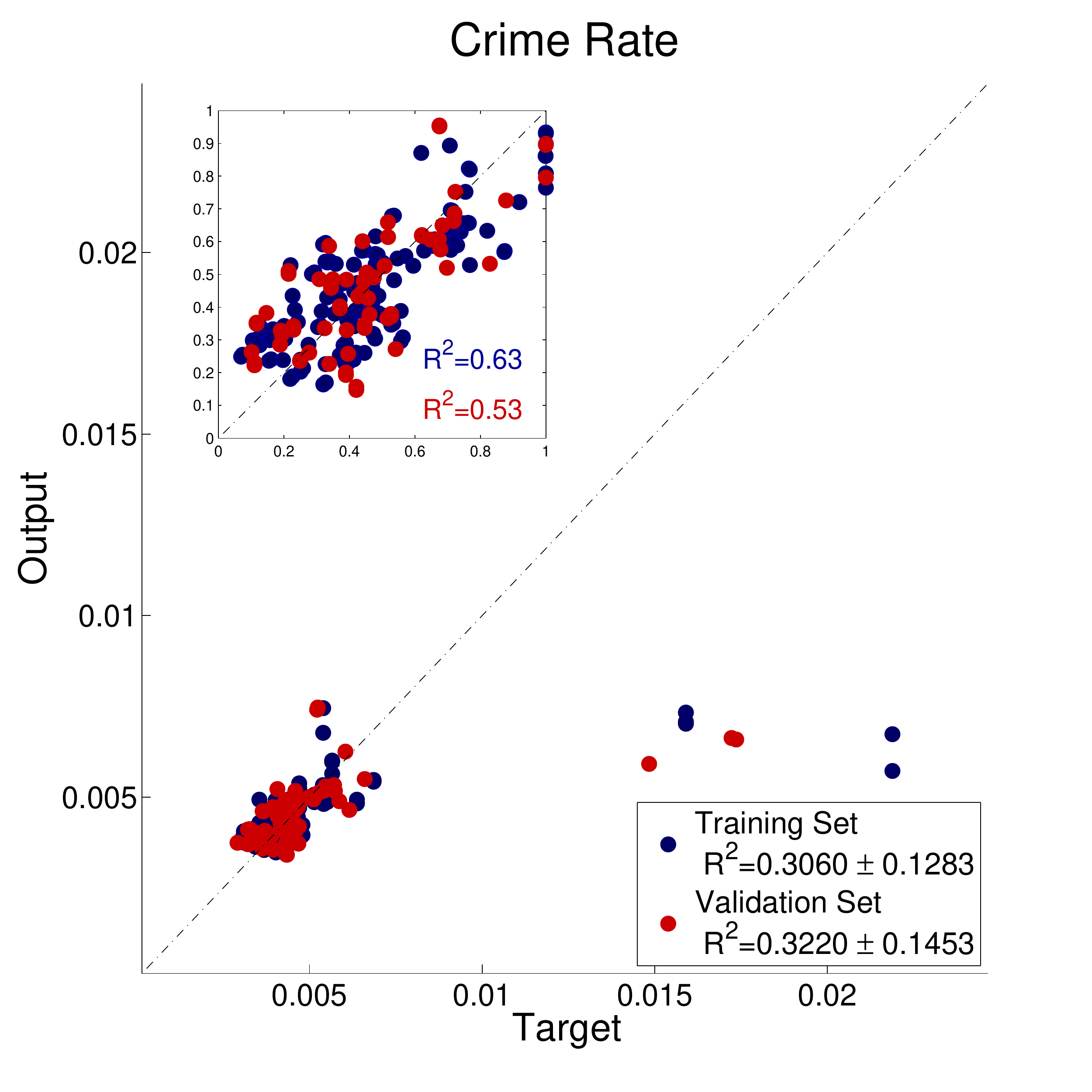}
                \label{fig:crime}
        \end{subfigure}%
        \begin{subfigure}[b]{0.4\textwidth}
                \includegraphics[width=\linewidth]{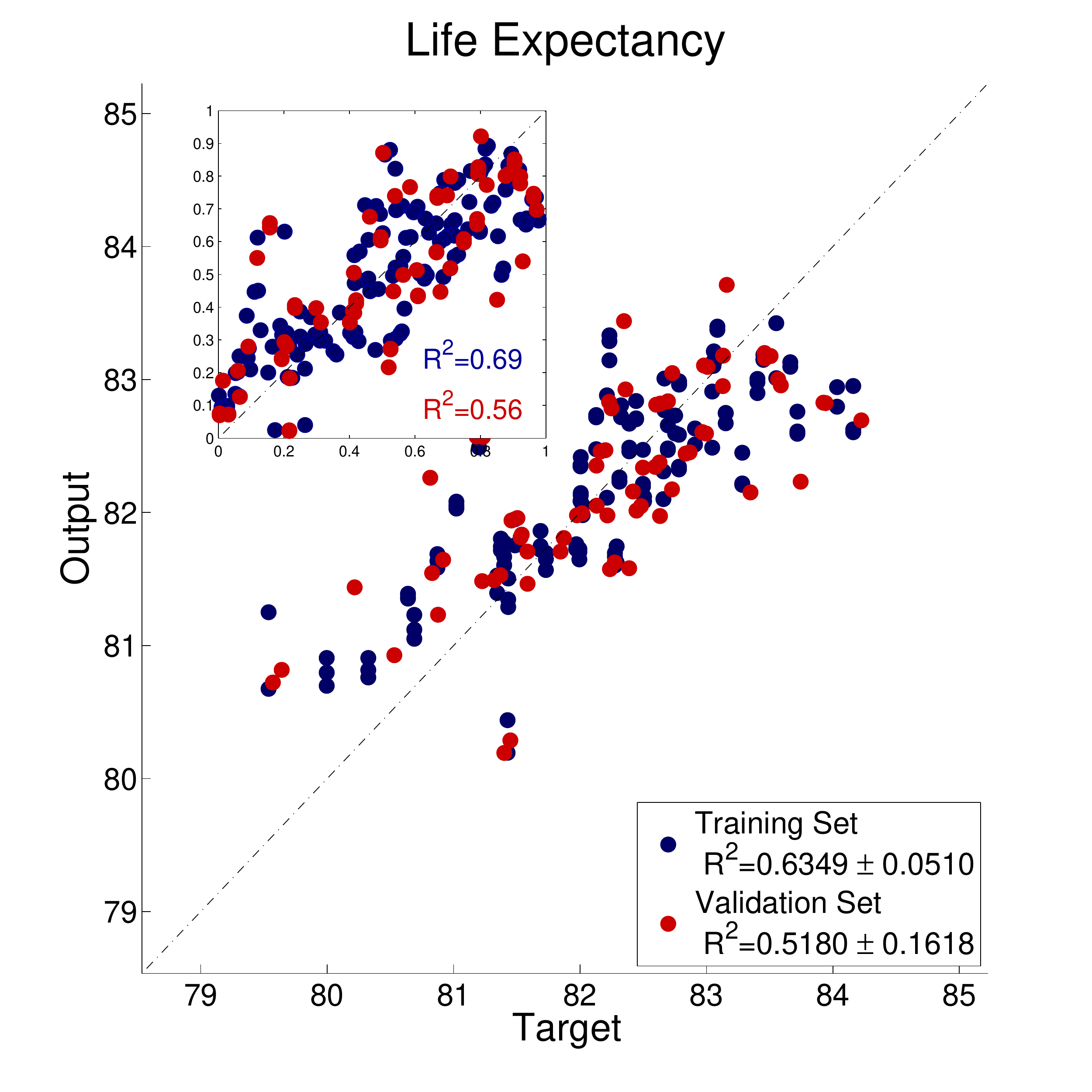}
                \label{fig:life}
        \end{subfigure}%
        \caption{Fitting/cross-validation results on both original and normalized (inset graphs). The $R^2$ is evaluated for both the training (blue) and validation (red) sessions.}
        \label{fig:validation}
\end{figure*}
\noindent
This way the normalized value of the predicted quantity $Y^{norm}$ is computed as:

\begin{equation}
Y^{norm} = g^{-1}(\mathbf{\beta}^T\mathbf{X}) = \frac{\exp(\mathbf{\beta}^T\mathbf{X})}{1+\exp(\mathbf{\beta}^T\mathbf{X})}.
\end{equation}

The model predicts values on the normalized scale, while for the original values we will have a final step of applying an inverse cumulative distribution function $F^{-1}$ for the distribution we fitted during the normalization step. This way the final model uses a superposition of $g^{-1}$ and $F^{-1}$:

\begin{equation}
Y = F^{-1}\left(g^{-1}(\mathbf{\beta}^T\mathbf{X}) \right)
\end{equation}

To determine the degree to which the model fits our data, we use the standard R-squared ($R^2$) metric for the linear regression model, i.e., measure based on unweighted residual sums of squares. The benchmark is the residual sum of squares in the intercept-only model, with fitted mean $\bar{y}$. There are several equivalent ways to express $R^2$ in the linear regression model, but their analogs for nonlinear models differ. In this paper we use the (unweighted) residual sum of squares yield as:

\begin{equation}
R^2 = 1 - \frac{\sum_{i=1}^N \left( y_i - \hat{\mu_i} \right)^2}{\sum_{i=1}^N \left( y_i - \bar{y_i} \right)^2},
\end{equation}

\noindent
where $\hat{\mu_i}$ is the predicted value by the machine learning algorithm and $y_i$ is the \emph{original} value.

\section{Results}
\label{res}

In this section we test our method in order to predict 6 different statistical socio-economic indices at the province scale. The input of the machine learning algorithm is the certain number of principal components evaluated by PCA. The information distribution among the principal components is illustrated in~\figurename~\ref{fig:pca}a. Although the first one already represents the $30\%$ of the total data variance, one needs $16$ components to cover at least $95\%$ of the information. We reported the correlations between those 16 leading principal components, built as linear combinations of the original 35 indicators, and the 6 socio-economic statistical indices in~Table~\ref{tab:correlation}. 

The strongest positive or negative correlations are highlighted by red bars in~\figurename~\ref{fig:correlation} where only the first five components show strong correlation with at least one of the quantities to be predicted, so might serve as a particularly valuable contribution for the model. In the rest of paper we will consider different numbers of principle components to train our models on and see which one is the optimal one based on the model performance on the training sets. Worth repeating is that the above PCA statistics was computed for the entire data sample containing all 52 provinces and in this section we use different training sets. Nevertheless, results in both cases are not substantially different. 

In order to evaluate the performance of our model to predict every statistical quantity, for the sake of reliability and avoiding random effects that can always occur with such a small data sample like the considered one, we use the following strategy, adopting the so-called \emph{random subsampling} or \emph{repeated shuffle and split} ideas \cite{maimon2010data}: we consider $4$ different learning sessions with different combinations of the training and validation sets and for each of those we train the model using $34$ (i.e., approx. $2/3$) of the entire $52$ provinces and then test it on the remaining $18$ (i.e., approx. $1/3$). 

In this framework in every training step, before applying PCA, we normalize the original 35 indicators fitting a \emph{normal} or \emph{lognormal} distribution. After training the model, we apply it to compute the predictions of the normalized values for the validation sample followed by applying the inverse cumulative distribution function according to the fitted distribution in order to map the predicted values back to the original scale. This means that we make 4 experiments and for each session we calculate $R^2$ value for the model performance on both~--- training and validation --- sets. The model performance is then characterized by the average values of $R^2$ on the validation sets. Moreover, we use 6 leading principle components as a feature set for our model.

As mentioned before, the results depend on how many leading principle components we consider. \figurename~\ref{fig:pca}b reported the dependence between the average values of $R^2$ for training and validation sets with this number. The optimal number of principle components one can pick up based on the model's performance on the training set. In our case it is $6$. Namely, performance goes up while we add new components until the 6th one and then adding more of them rather than contributing, it starts to affect the results negatively. Six principle components also give the best performance on the validation sets. A spatial visualization presenting our model performance on the example of GDP is given in Figure \ref{fig:gdp_map}, showing quite good general match besides of couple of specific outliers.

The quantitative analysis of the model performance on both original and normalized scales for our statistical indices is presented in \figurename~\ref{fig:validation}. We got very good results in predicting all the considered quantities, except for the crime rate parameter in which case our model is not able to closely predict some specific outliers on the original value scale and this consequently affects the corresponding $R^2$ score. The score for the normalized scale is already much better (over $50\%$) showing that this issue is only the matter of scale. For all the rest, $R^2$ values for the validation set range around $50-60\%$, while for the training set between $60-80\%$. To conclusion, overall the R-squared coefficient (i.e., $\%$ of parameter variation explained by the model) is slightly lower for the validation sample (red points in \figurename~\ref{fig:validation}) compared to the training sample. Nevertheless, the difference is rather small showing that our approach of training the model made it possible to largely avoid overfitting.

\section{Conclusions}
\label{conclusions}

In this paper we proposed 35 different characteristics of individual economic behavior quantifiable through the dataset of anonymized bank card transactions, and then evaluated them on the example of Spain. We showed that those quantities could be used for estimating economic performance of the regions in the country, as proposed supervised machine learning technique demonstrated to perform well on the validation samples for predicting major official statistical quantities such as GDP, housing prices, unemployment rate, level of higher education, life expectancy and crime rate on the level of Spanish provinces. 

Moreover, the same approach is applicable in cases when official statistics is not available or is inconsistent, for example when considering geographical units of a finer spatial scale such as municipalities, districts or neighborhoods. The approach also allows evaluating temporal variation of economic performance of the regions, which is especially useful since official statistics is more static and cannot give a really fine-grained longitudinal perspective. Finally, the proposed model can be further employed for estimating more specific characteristics of local economic performance addressing particular business needs.

\section*{Acknowledgement}

The authors would like to thank Banco Bilbao Vizcaya Argentaria (BBVA) for providing the anonymized bank dataset. Special thanks to Behrooz Hashemian for his valuable feedback and suggestion on the analysis methodology as well as to Assaf Biderman, Marco Bressan, Elena Alfaro Martinez and Maria Hernandez Rubio for organizational support of the project and stimulating discussions. We further thank BBVA, MIT SMART Program, Accenture, Air Liquide, The Coca Cola Company, Emirates Integrated Telecommunications Company, The ENEL foundation, Ericsson, Expo 2015, Ferrovial, Liberty Mutual, The Regional Municipality of Wood Buffalo, Volkswagen Electronics Research Lab, UBER and all the members of the MIT Senseable City Lab Consortium for supporting the research. Finally, the authors also acknowledge support of the research project "Managing Trust and Coordinating Interactions in Smart Networks of People, Machines and Organizations", funded by the Croatian Science Foundation.

\bibliographystyle{ieeetr}
\bibliography{bibliography}

\bigskip

\end{document}